\documentstyle[12pt,psfig]{article}

\def\baselinestretch{1.2}
\def\href#1#2{#2}  

\newcommand{\norm}[1]{\raise.3ex\hbox{:} #1 \raise.3ex\hbox{:}\,}

\newcommand{\beq}{\begin{equation}}
\newcommand{\eeq}{\end{equation}}

\newcommand{\beqar}{\begin{eqnarray}}
\newcommand{\eeqar}{\end{eqnarray}}

\def\appendix{{\newpage\section*{Appendix}}\let\appendix\section%
        {\setcounter{section}{0}
        \gdef\thesection{\Alph{section}}}\section}

\catcode`\@=11

\@addtoreset{equation}{section}
\catcode`\@=12

\begin{document}

\begin{titlepage}

\begin{flushright}
NSF-ITP-98-127\\
hep-th/9812159
\end{flushright}
\vfil\vfil

\begin{center}

{\Large {\bf Supergravity Solutions for Localized Intersections of Branes}}

\vfil

Akikazu Hashimoto

\vfil

Institute for Theoretical Physics\\
 University of California\\
 Santa Barbara, CA  93106

\end{center}

\vspace{5mm}

\begin{abstract}
\noindent 
We construct an explicit supergravity solution for a configuration of
localized D4-brane ending on a D6-brane, restricted to the near
horizon region of the latter. We generate this solution by
dimensionally reducing the supergravity solution for a flat M5-brane
in $R^{1,7} \times C^2/Z_N$ with the M5-brane partially embedded in
$C^2/Z_N$. We describe the general class of localized intersections and
overlaps whose supergravity solutions are constructible in this way.
\end{abstract}

\vfil\vfil\vfil
\begin{flushleft}
December 1998
\end{flushleft}
\end{titlepage}

\renewcommand{\baselinestretch}{1.05}  

\section{Introduction}

Much of the recent advances in string theory are derived from the
properties of brane objects which exist in the theory.  Originally,
these objects were formulated as a $p$-brane solution \cite{HS91} to
the supergravity field equations which are the low-energy effective
description of string theory. The solutions described in \cite{HS91}
were flat along the world-volume and spherically symmetric in the
directions transverse to the brane. A possibility to study these
objects microscopically was opened following the realization that
certain class of these branes are D-branes, a type of topological
defect on which a string can end \cite{Dnotes}. Even branes which are
not D-branes can in general be thought of as their duals.

From the perspective of string theory, it is easy and instructive to
consider a more complicated configuration such as localized (as
opposed to smeared) intersections and junctions of strings. Wide range
of applications of these brane configurations have been discussed by
many authors, including such topics as the brane construction of gauge
theories \cite{Giveon:1998sr}, BIons \cite{CM97,Gibbons97}, string
junctions and networks \cite{AHK97,DM97,Sen97}, just to name a few.
Related configurations for topological defects in ordinary field
theories have been discussed in \cite{Carroll:1998pz}.

It is natural to wonder the extent to which phenomena associated with
these non-trivial brane configurations could have been discussed in
the original formulation of branes as solutions to the supergravity
equations of motion.  This question is of acute interest especially in
light of the recent realization that supergravity provides an
effective dual description to strongly coupled large $N$ gauge
dynamics \cite{juanAds}.  Unfortunately, progress in this line of
investigation is obstructed by our lack of knowledge of the explicit
supergravity solutions corresponding to these backgrounds.  Localized
intersections break the spherical symmetry along the the transverse
directions, and techniques for solving supergravity field equations in
such a general context is not yet known. Indeed, explicit form of the
solutions are known only for the few special cases
\cite{Tseytlin:1996as,Tseytlin:1997cs,Khuri:1993cs,Gauntlett:1996pb,Gauntlett:1997pk}.

Recently, interest have focused on the near horizon geometry of branes
following the work of \cite{juanAds}.  By restricting one's attention
to the near horizon geometry, the problem of finding supergravity
background corresponding to localized intersections also simplify.
Explicit expression for D-instanton localized in the near horizon
AdS$_5$ geometry of the D3-brane and its relation to gauge theory
instantons were discussed in \cite{Chu:1998in,Balasubramanian:1998de}.

In a related but independent line of development, the authors of
\cite{Itzhaki:1998uz} have shown that a certain localized brane
solution can be derived explicitly in the near horizon region of the
D6-brane. The main observation behind \cite{Itzhaki:1998uz} is the
fact that the near horizon geometry of $N$ D6-brane in type IIA
supergravity admits an interpretation as a dimensional reduction of
11-dimensional supergravity on $R^{1,6} \times \Sigma$ where $\Sigma$
is an ALE space $C^2 / Z_N$ \cite{Itzhaki:1998dd}. It's universal
cover is therefore the flat 11-dimensional Minkowski space.  The
authors of \cite{Itzhaki:1998uz} then considers the supergravity
solution of the flat M2 and M5 branes embedded in the $R^{1,6}$ part
of the universal cover.  By moding the action of the $Z_N$ group and
reducing to type IIA, they constructed localized supergravity
solutions of D2 and NS5-branes in the near horizon region of the
D6-brane, as well as other configurations related by U-duality.

The main goal of this paper is to point out that it is possible to
partially (or entirely) embed the M2 and M5 into the ALE part of the
space in such a way that these branes are flat in the universal cover.
Therefore, by following the same steps of orbifolding and
dimensionally reducing, one can obtain an explicit expression for a
new class of localized brane geometry in the near horizon region of
the D6-brane.  Although these solutions are far from being general,
such an explicit solution might serve as a useful aid in the on-going
pursuit for a general construction of supergravity solutions for
localized intersections.

This paper is organized as follows. We begin in section 2 by reviewing
the basic construction of the near horizon D6-brane geometry as a dimensional
reduction of an ALE orbifold, and present an explicit construction for
the space-time background of a localized configuration of D4-brane
ending on a near-horizon core of the D6-brane.  In section 3, we
explore a general class of brane configurations which can be
obtained using this method and present few more examples, including
the overlapping NS5 and D6-branes. Comments and conclusions are
collected in section 4.

\section{D4-brane ending on D6-brane}

We begin this section by briefly reviewing the basic construction of
the near horizon geometry of the D6-brane as a dimensional reduction
of an ALE orbifold.  This was essentially worked out in
\cite{Itzhaki:1998uz,Itzhaki:1998dd} but we repeat it here for
completeness and to fix the notation.

Let us begin by considering 11-dimensional supergravity on $R^{1,6}
\times \Sigma$ where $\Sigma$ is an ALE space of type $A_{N-1}$. For
our purposes, it is convenient to think of this space as a $Z_N$
orbifold of $(v,w) \in C^2$ under the action $(v,w) \rightarrow (e^{2
\pi i / N} v,e^{2 \pi i / N} w)$. By setting $v = x_8 + i x_9$ and $w =
x_7 + i x_{10}$, we can think of $R^{1,6} \times \Sigma$ as
$R^{1,10}/Z_N$ with $Z_N$ acting on (7,8,9,10) coordinates.

The metric on $R^{1,10}$ is simply
$$ds_{11}^2 = -dx_0^2 + \sum_{i=1}^{10} dx_i^2.$$
To make the $Z_N$ action on this background transparent, it is convenient to change variables
\beq
w = x_7 + i x_{10} = \rho e^{i \tilde \phi} \cos(\tilde\theta), \qquad
v = x_8 + i x_{9} = \rho e^{i \tilde \psi} \sin(\tilde\theta)
\label{cvar1}
\eeq with ranges $0 \le \tilde \theta \le \pi/2$, $0 \le \tilde \phi,
\tilde \psi \le 2 \pi$, and $0 \le \rho$. Orbifolding by $Z_N$
instructs us to identify $(\tilde \phi, \tilde \psi) \sim (\tilde
\phi, \tilde \psi) + (2 \pi/N, 2 \pi/N)$. In terms of these variables,
the metric takes the form
$$ ds_{11}^2 = dx_\parallel^2 + d \rho^2 + \rho^2(d \tilde \theta^2 + \sin^2(\tilde \theta) d \tilde \psi^2 + \cos^2(\tilde \theta) d \tilde \phi^2).$$
Further change of variables
\beq
U = {\rho^2 \over 2 N l_p^3}, \quad \theta = 2 \tilde \theta, \quad
\psi = \tilde\psi - \tilde\phi, \quad x_{11} = R_{11} \phi = R_{11} N
\tilde \phi
\label{cvar2}
\eeq
leads to a metric of the form
$$ds_{11}^2 = dx_\parallel^2  +{l_p^3 N \over 2U} dU^2 + {l_p^3 N U \over
2} (d \theta + \sin^2(\theta) d \psi) + {2 U l_p^3 \over N R_{11}} [ d
x_{11} + {N R_{11} \over 2} (\cos(\theta)-1) d \psi]^2$$
where $x_{11}$ is periodic under shift by $2 \pi
R_{11}$. Dimensionally reducing along $x_{11}$ leads to a type IIA
metric of the form
\begin{eqnarray*}
ds_{10}^2 & = & \alpha' \left[ {(2 \pi)^2 \over g_{YM}} \sqrt{{2 U
\over N}} dx_{\parallel}^2 + {g_{YM} \over (2\pi)^2} \sqrt{{N \over
2U}} dU^2 + {g_{YM} \over (2 \pi)^2 \sqrt{2}} \sqrt{N} U^{3/2}
d\Omega_2^2 \right], \\
e^\phi & = & {g_{YM}^2 \over 2 \pi} \left( {2 U \over g_{YM}^2 N
}\right)^{3/4}, \\
A^{(1)} & = & {N R_{11} \over 2} (\cos(\theta)-1) d \psi,
\end{eqnarray*}
where $d\Omega_2^2 = r^2 (d \theta^2 + \sin^2 (\theta) d
\psi^2)$. This is precisely the type IIA supergravity background for
$N$ D6-branes \cite{HS91} in the near horizon limit
\cite{Itzhaki:1998dd}
\beq
U = {|x| \over \alpha'} = {\rm fixed}, \quad g_{YM}^2 = (2 \pi)^{p-2} g_{st} l_s^{p-3} = {\rm fixed},\quad  p=6, \quad \alpha' \rightarrow 0.
\label{nhlimit}
\eeq

The authors of \cite{Itzhaki:1998uz} pointed out that a similar
procedure can be carried out starting with a background of $n$
M2-brane extended along the (0,1,2) direction and embedded anywhere in
the (3,4,5,6) direction and into the origin of the (7,8,9,10)
coordinates of $R^{1,10}$. The supergravity solution for such a
background takes a simple form
$$ds_{11}^2  =  f^{-2/3}( -dx_0^2 + dx_1^2 + dx_2^2 ) + f^{1/3}(dx_3^2 + dx_4^2 + dx_5^2 + dx_6^2 + dx_7^2+dx_8^2+dx_9^2 + dx_{10}^2)$$
$$A^{(3)}  =  {f-1 \over f} dx_0 \wedge dx_1 \wedge dx_2$$
and is invariant under the action of $Z_N$.  Here, $f$ is the standard
harmonic function for the M2-brane. One can therefore orbifold and
dimensionally reduce along the same coordinates as in the pure ALE
case. Since the dimensional reduction maps M2 brane into D2 brane, one
obtains by this construction a supergravity solution for D2 brane
localized along (3,4,5,6) coordinates in the near horizon geometry of
the D6-brane. Similar procedure using the M5 brane leads to a
construction of NS5-brane localized within D6-branes.

The main point of this paper is to show that a new class of brane
configuration can be generated using the same technique as
\cite{Itzhaki:1998uz} but letting M-branes partially (or entirely)
wrap $\Sigma$. As long as such embeddings arise from orbifolding a
configuration of flat M-branes in $R^{1,10}$, one can orbifold and
dimensionally reduce the M-brane supergravity solution.  In the
remainder of this section, we will discuss one concrete example in
detail, the configuration of D4-brane ending on a D6-brane.
Generalizations will be discussed in the following section.

D4-branes arise from double dimensional reduction of M5-brane, so let
us consider a configuration of M5-brane in $R^{1,10}$ oriented along
the (0,1,2,3,7,10) direction, localized anywhere along the (4,5,6)
direction, and localized at the origin along the (8,9) direction.  The
$Z_N$ orbifold group acts on complex coordinate $w = x_7 + i x_{10}$
and $v = x_8 + i x_9$, so such a configuration is invariant under the
action of $Z_N$. The idea is simply to take the metric of M5-brane in
$R^{1,10}$
$$ds_{11} = f^{-1/3} (-dx_0^2 + dx_1^2 + dx_2^2 + dx_3^2 + dx_7^2 + dx_{10}^2) + f^{2/3} (dx_4^2 + dx_5^2 + dx_6^2 + dx_8^2 + dx_9^2)$$
$$A_6 = {f-1 \over f} dx_0 \wedge dx_1 \wedge dx_2 \wedge dx_3 \wedge dx_7 \wedge dx_{10} $$
where $A^{(6)}$ is the Poincare dual of the 3-form of the
11-dimensional supergravity, and perform the same manipulation
performed in the above. Since all this is quite straightforward, we
simply quote the result here:
\begin{eqnarray}
ds^2 & = & \alpha' \left[ {(2 \pi)^2 \over g_{YM}} \sqrt{{2 U \over
N}} \left( \Delta^{1/2}\left(\sum_{i=0}^3 dx_i^2 + \sum_{i=4}^6 f
dx_i^2\right) \right) + {g_{YM} \over (2 \pi)^2} \sqrt{{N \over 2 U}}
f \Delta^{3/2} dU^2\right. \nonumber \\
&&  \left.+\quad {g_{YM} \over (2 \pi)^2} \sqrt{N \over 2} U^{3/2}
\left( f \Delta^{1/2} \tilde{\Delta} d\theta^2 + \Delta^{-1/2}
\sin(\theta)^2 d \psi^2 + {f-1 \over U} \Delta^{1/2} \sin(\theta) dU
d\theta \right) \right] \nonumber \\
A^{(1)} & = & {\Delta^{-1} N R_{11}\over 2}(1-\cos(\theta)) d\psi \nonumber \\
A^{(5)} & = & \left({f - 1 \over f}\right) dx_0 \wedge dx_1 \wedge
 dx_2 \wedge dx_3 \wedge \left({l_p^3 (1+\cos(\theta)) \over 4 R_{11}}
 dU - {l_p^3 U \sin(\theta) \over 4 R_{11}} d\theta \right) \nonumber
 \\
e^{\phi} & = & \sqrt{f} \left({g_{YM}^2 \Delta^3 U^3 \over 2N^3
\pi^4}\right)^{1/4}
\end{eqnarray}
where
$$\Delta = {1 + f +  (1 - f) \cos(\theta) \over 2 f}, \qquad
\tilde\Delta = {1 + f -  (1 - f) \cos(\theta) \over 2 f},$$
$$ f = 1 + {\pi n l_p^3 \over r^3}, \qquad r^2 = \sum_{i=4}^6 x_i^2 + 2 N l_p^3 U \sin^2(\theta/2).$$
To derive  $A^{(5)}$, we used the fact that
\begin{equation}
dx_i = {dx_i \over dU} dU + {dx_i \over d\theta} d\theta + {dx_i \over d\psi} d\psi + {dx_i \over dx_{11}} dx_{11}
\end{equation}
This, therefore, is the metric for the localized intersection of a
D4-brane and a D6-brane.  A useful check is to see that the above
expression simplifies to that of pure D6-brane near horizon geometry
upon setting $f=1$. Note that the D4-brane is ending on instead of
piercing through the D6-brane, as can be seen from the geometry of the
$r=0$ locus. To emphasize the point that this is a classical
supergravity solution, all dependence on $l_s$, $l_p$, and $g_{st}$ can
be scaled into (recalling $g_{YM}^2 = (2 \pi)^{p-2} g_{st} l_s^{p-3}$)
$$y = \alpha' U, \quad Y = { g_{st} N l_s \over 2}, \quad R^3 = \pi n \l_p^3.$$
In these variables, the background becomes
\begin{eqnarray}
ds^2 & = & \sqrt{{y \over Y}} \left( \Delta^{1/2} \left(\sum_{i=0}^3
dx_i^2 + \sum_{i=4}^6 f dx_i^2\right) \right) + \sqrt{{Y \over y}} f
\Delta^{3/2} dy^2 \nonumber \\
&& \quad y^{3/2} \sqrt{Y} \left( f \Delta^{1/2} \tilde{\Delta}
d\theta^2 + \Delta^{-1/2} \sin(\theta)^2 d \psi^2 + {(f-1) \over y}
\Delta^{1/2} \sin(\theta)dy d\theta \right) \nonumber \\
A^{(1)} & = & \Delta^{-1} Y(1-\cos(\theta)) d\psi \nonumber \\
A^{(5)} & = & \left({f - 1 \over f}\right) dx_0 \wedge dx_1 \wedge
dx_2 \wedge dx_3 \wedge \left({ 1+\cos(\theta) \over 4} dy
 - { y \sin(\theta) \over 4 } d\theta \right)  \nonumber \\
{1 \over g_{st}} e^{\phi} & = &  \sqrt{f} \left({y  \over Y} \Delta \right)^{3/4} 
\end{eqnarray}
with
$$ f = 1 + {R^3 \over r^3}, \qquad r^2 = \sum_{i=4}^6 x_i^2 + 4 y Y \sin^2(\theta/2).$$

\section{Some generalizations}

In the previous section, we discussed one concrete example of
spacetime background generated by orbifolding and dimensionally
reducing a configuration of flat M-brane in $R^{1,10}$.  In this
section, we will describe the scope to which this construction can be
generalized, and present some examples. In M-theory, we have M2 and
M5-branes to serve as our basic building blocks. One can in principle
wrap any part of the M2 or M5-brane world volume in the $R^{1,6}$ part
of the target space and all the rest in the $\Sigma$. The embedding
into $\Sigma$ must be such that enough supersymmetry is preserved to
ensure the stability of the configuration.  The issue of supersymmetry
is easier to address in the case where the number of dimensions
embedded into $\Sigma$ is even, where the allowed embedding geometries
are holomorphic curves. The possibility therefore is to embed none,
two, or four of the world volume coordinates of the M-branes into
$\Sigma$.  Embedding none of the coordinates was the case discussed in
\cite{Itzhaki:1998dd}. Only M5-brane can be used to embed four of the
world volume coordinates into $\Sigma$, and this leads simply to a
configuration of D4-brane and D6-brane sharing one space and one time
coordinates, and relatively transverse in the remaining eight
coordinates. The example described in the previous section falls in
the category of embedding two of the world volume coordinates into
$\Sigma$. In the remainder of this section, we will elaborate on this
possibility and present further examples of supergravity backgrounds
for brane intersections. We will focus mainly on embedding M5-branes
although it is straightforward to extend these results to M2-branes.

We are interested in finding a holomorphic embedding of complex plane
into $\Sigma$.  Similar holomorphic embedding of two the world volume
coordinates of the M5 brane into a Taub-NUT space was discussed in
\cite{Nakatsu:1997ty}. These authors studied the M-theory origin of
the Hanany-Witten transition between D6 and NS5-branes
\cite{Elitzur:1997fh} by considering a generic holomorphic embeddings
in Taub-NUT space and studying its limiting behavior as the radius of
11-th circle at infinity is sent to zero. For our discussion of
constructing supergravity solutions, however, we must restrict ourself
to the near horizon ALE geometry, since the fact that the M-theory
lift is an orbifold of flat space is crucial to our construction.
Furthermore, we are restricted to flat embeddings.

The flat curve describing the embedding of the previous section was
simply $v=0$. In general, flat embedding in $C^2$ is given by a linear
curve $v - aw = b$, where $a$ and $b$ are complex parameters. We are
further restricted by the requirement that our brane configuration be
invariant under the action of the orbifold group which sends $(v,w)$
to $(e^{2 \pi i/N} v,e^{2 \pi i/N} w)$. This requirement can be
satisfied by adding images, or equivalently, by taking
\beq
(v-aw)^N = b^N
\label{curve}
\eeq
as our embedding curve.  In the remainder of this section we will
describe the geometry and the supergravity background for this brane
configuration.

\subsection{Geometry of the brane}

\begin{figure}
\centerline{\psfig{file=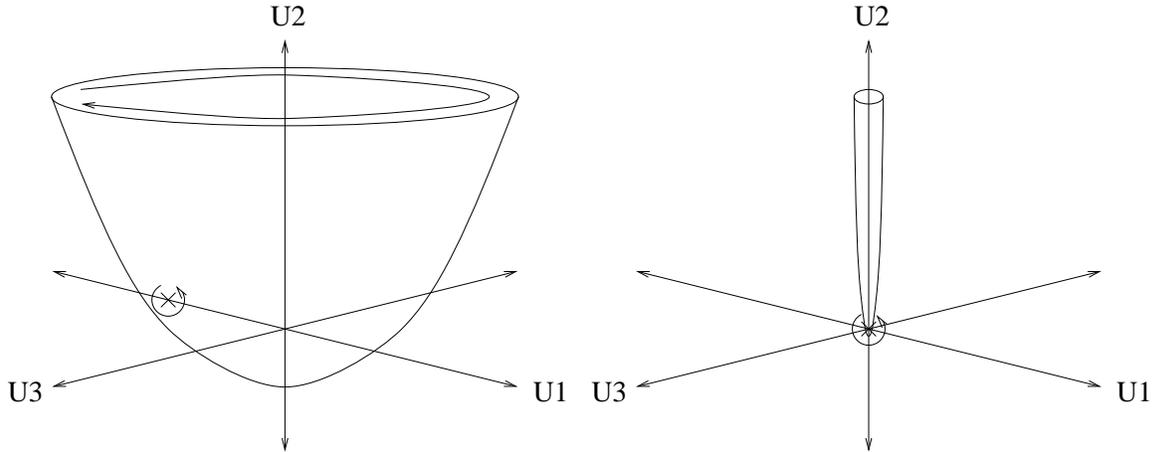,width=6in,angle=270}}
\caption{Configuration of branes embedded according to $(v-aw)^N =
b^N$ in the natural coordinates of the near horizon D6-brane
geometry. Here, $a=1$ indicated by the fact that the paraboloid points
along the $U_2$ direction.  As $b \rightarrow 0$, the paraboloid
degenerates to a ``string'' like geometry. \label{figa}}
\end{figure}

Let us first describe the geometry of this brane configuration from
the point of view of type IIA near horizon D6-brane point of view.
The embedding (\ref{curve}) turns out to describe a paraboloid
oriented along the axis
$$a = \tan(\theta/2) e ^{i \psi}$$
with the focus located at the origin.  This is demonstrated by taking
the norm of (\ref{curve}) 
\beq
|v - aw | = |b|
\label{norm}
\eeq
and applying a chain of change of variables. Let us assume without
loss in generality that $a$ is real.  The specific change of variables
are as follows. First apply (\ref{cvar1}) and (\ref{cvar2}) to express
(\ref{norm}) in terms variables $U$, $\theta$, $\psi$, and
$x_{11}$. Then go to Cartesian coordinates by defining
$$U_1 = U \cos(\theta), \quad U_2 + i U_3 = U \sin(\theta) e^{i \psi}.$$
Finally, rotate the coordinate system by defining
$$ \tilde{U}_1 = {a^2-1 \over 1+a^2} U_1 + {2 a \over 1+a^2} U_2, \quad \tilde{U}_2 = -{2a \over 1+a^2} U_1 + {a^2-1 \over 1+a^2} U_2, \quad \tilde{U}_3 = U_3.$$
In terms of these new variables (\ref{norm}) becomes
$$\tilde{U}_1 =  - {\frac{{b^2}}{2 \left( 1 + {a^2} \right)  {{l_p}^3} N}} + 
   {\frac{\left( 1 + {a^2} \right)  {{l_p}^3} N }
     {2 {b^2}}} (\tilde{U}_2^2 + \tilde{U}_3^2)
$$
which shows that this is indeed a paraboloid. We are also interested
in how this paraboloid is embedded in the $x_{11}$ coordinate. A
convenient way to address this issue is to parameterize the paraboloid
by $z = v/w$ which amounts to labeling the points on the paraboloid by
its intersection with a ray from the origin oriented along $z =
\tan(\theta/2) e^{i \psi}$. In terms of $z$, (\ref{curve}) is simply
$$w^N = {b^N \over (z-a)^N}.$$
The analytic structure of this embedding is now very clear. There is a
pole at $z=a$ and a zero at $z=\infty$. This means that $w$ winds once
around the complex plain if one winds $z$ around $a$, and $w$ winds
once in the opposite direction if one winds $z$ around $\infty$. The
points $z=a$ and $z=\infty$ corresponds to the point at infinity and
$\theta = \pi$, respectively, on the paraboloid.  Since winding once
in $w$ corresponds to winding $N$ times around $x_{11}$, we find that
the paraboloid winds $N$ times around circle of 11-th dimension around
these two points (See figure \ref{figa}).

The effects of adjusting parameters $a$ and $b$ are easy to
understand. The parameter $a$ controls the orientation of the
paraboloid whereas the parameter $b$ controls the eccentricity.  For a
generic value of $a$, the parabola degenerates into a thin tube as we
send $b$ to zero. In this limit, the point at $\theta = \pi$ also
approaches the origin, and we recover the configuration of D4-brane
ending on the core of D6-brane as in the previous section. The case
where $a = \infty$ (keeping  $b/a$ finite) appears to be an
exception of this general behavior. In this case, the axis of the
parabola also points along $\theta = \pi$ direction. This will cause
the zero and the pole in the $z$ plane to merge, resulting in a
configuration that does not wrap the 11-th direction. Taking $b/a$ to
zero will also degenerate the paraboloid into a tube, but it will not
wind around the 11-th direction and should not be thought of as a
D4-brane. Instead, it a highly curved configuration of NS5-brane
stabilized by the gravitational background of the D6-brane.

What we are finding is the point $b=0$ and $a=\infty$ is a special
point in the moduli-space of curves $(v-aw)^N = b^N$. The moduli-space
of brane configurations with different topology of winding around the
11-th direction is meeting at this point. This structure is strongly
reminiscent of the Hanany-Witten transition
\cite{Giveon:1998sr,Elitzur:1997fh}.

\subsection{Supergravity Solution}

Now that we understand the geometry associated with this embedding,
let us discuss the corresponding supergravity solution. Although the
procedure is very straightforward extension of the result of
\cite{Itzhaki:1998uz} and the previous section, the explicit results
are rather lengthy and cumbersome. We will therefore present only
schematic results in this section. Enough details will be present so
that explicit results, if desired, can easily be reproduced.

The curve (\ref{curve}) describes $N$ parallel M5-branes in
$R^{1,10}$.  This will naturally separate the $(v,w)$ plane into
components parallel and transverse to these branes:
\beq z_\parallel = {1 \over \sqrt{1+|a|^2} }(w + \bar a v), \qquad
z_\perp = {1 \over \sqrt{1+|a|^2} } (v - a w) \eeq
The supergravity solution describing these $N$ sets of M5-branes is
\begin{eqnarray}
ds_{11}^2 &=& f^{-1/3}(-dx_0^2 + dx_1^2 + dx_2^2 + dx_3^2 +
dz_\parallel d \bar{z}_\parallel) + f^{2/3} (dx_4^2 + dx_5^2 + dx_6^2
+ dz_\perp d\bar z_\perp) \nonumber \\
A^{(6)}& =& {f -1 \over f} dx_0 \wedge dx_1 \wedge dx_2 \wedge dx_3 \wedge \left({dz_\parallel \wedge d \bar{z}_\parallel \over 2i}\right) \label{newsol}
\end{eqnarray}
where the harmonic function $f$ encodes the fact that there are $N$ sets
of M5-branes
\beq f = 1 + \sum_k {\pi n l_p^3 \over r_k^3}, \qquad r_k^2 = x_4^2 +
x_5^2 + x_6^2 + |z- b e^{2 \pi i k/N}|^2. \label{harm} \eeq

All that remains to be done is to follow the change of variables
(\ref{cvar1}) and (\ref{cvar2}) and to dimensionally reduce. The only
subtlety that arises for this general case is the fact that the
harmonic function $f$ can have explicit dependence on $x_{11}$. This
means that background massive Kaluza-Klein fields are generated when
(\ref{newsol}) is dimensionally reduced. These massive Kaluza-Klein
excitations can be eliminated if, instead of localizing the M5-branes
in the original $R^{1,10}$ space along a polygon, we smear along the
circle of radius $|b|$. This will change the form of the harmonic
function from (\ref{harm}) to
\beq f = 1 + N \int_0^1 dq {\pi n l_p^3 \over ( x_4^2 + x_5^2 + x_6^2 + |z- b
e^{2 \pi i q}|^2)^{3/2}}. \label{harm2} \label{dz} \eeq
Of course, to leading order in large $N$, the background is already
smeared and this extra step is not necessary.

Since these backgrounds describe coexisting D4-branes and NS5-branes
in the near horizon region of the D6-brane, it would be interesting to
examine the behavior of RR 5-form and NSNS 6-form (Poincare dual of
NSNS 2-form) in this background. Here, we seem to encounter a small
puzzle regarding these backgrounds. Let us illustrate this with some
examples.

First, consider sending $a$ to zero so that we have a curve $v^N =
b^N$. If we also set $b$ to zero, this reduces to $N$-fold copy of the
situation discussed in the previous section.  There, we found that
only the RR 5-form background and no NSNS 6-form background was
generated, which is appropriate for a background consisting only of
D4-branes and D6-branes. The fact that only RR 5-form follows from the
fact that the M-theory 6-form (\ref{newsol}) contains a factor
\beq
\left({dz_\parallel \wedge d \bar{z}_\parallel \over 2i}\right)
=dx_7 \wedge dx_{10} = {l_p^3 (1+\cos(\theta)) \over 4 R_{11}} dU
\wedge dx_{11} - {l_p^3 U \sin(\theta) \over 4 R_{11}} d\theta \wedge
dx_{11},
\label{form1}
\eeq
which is to say that the factor of $dx_{11}$ is always
present. Therefore, this 6-form was guaranteed to reduce to a 5-form
in type IIA.

Although this was perfectly fine for $v^N =0$, this is somewhat
troublesome for $v^N=b^N$. For by increasing $b$ to non-zero values,
what used to be a string-like object (tensored with $R^{1,3}$) blows
up into a paraboloid (times $R^{1,3}$). It would seem such an object
is better described as NS5-brane, and we would expect NSNS 6-form to
turn on in this background.

On the other hand, $dz_{\parallel}$ does not change under change in
$b$ (\ref{dz}).  Only the form of the harmonic function $f$ changes
with changes in $b$. Therefore, the 6-form in 11 dimensions will
always have a $dx_{11}$ component, and the 6-form will always reduce
to RR 5-form in type IIA.

Similar paradox appears for other configurations. For example, if we
consider the curve in the $a \rightarrow \infty$ limit $w^N = b^N$,
where the brane is no longer wrapping the circle in the 11-th
dimension, we expect not to have RR 5-form in the background. It can
be easily checked, however, that
\beq
dx_8 \wedge dx_{9} = {l_p^3 (1-\cos(\theta)) \over 4 R_{11}} dU
\wedge (dx_{11} + N R_{11} d \psi) + {l_p^3 U \sin(\theta) \over 4
R_{11}} d\theta \wedge (dx_{11} + N R_{11} d \psi).
\label{form2}
\eeq
This time, the M-theory 6-form has components containing and not
containing the factor of $dx_{11}$. This means that when reduced to
type IIA, both RR 5-form and NSNS 6-form is excited.

Finally, consider the embedding $v - aw = 0$. We showed that this
embedding describes a parabola that collapsed down to a thin tube,
and should be thought of as a D4-brane stretching along $a =
\tan(\theta/2) e^{i \psi}$ and ending on the D6-brane core. However,
${1 \over 2i} dx_\parallel \wedge d \bar{z}_\parallel$ will be some
linear combination of (\ref{form1}) and (\ref{form2}), indicating that
both NSNS 6-form and the RR 5-form background is turned on.

It is not clear if this apparent mismatch between the ``identity'' of
the branes and the differential forms in the background is really a
problem.  Perhaps there is no clear distinction between NS5-branes and
D4-branes in the near horizon region of a D6-brane.

\section{Discussions}

The main goal of this paper was to demonstrate that a certain
supergravity solution for locally intersecting branes can be derived
by taking a flat configuration of M-brane in $R^{1,10}$ and acting by
quotient group and dimensionally reducing to type IIA.  As a concrete
example, we derived the supergravity background for D4-brane ending on
a D6-brane, albeit only in the near horizon limit of the D6-brane
background. This configuration is precisely that of a D4-brane BIon
ending on the D6-brane.  Unfortunately, the near horizon geometry of
the D6-brane does not in any way correspond to a ``decoupling limit''
of some world volume field theory
\cite{Itzhaki:1998dd,Seiberg:1997ad}, so the supergravity solution we
derived in section 2 are not expected to capture any aspect of
magnetic monopole solution of 6+1-dimensional gauge theory (whatever
that may be). Nonetheless, for small $g_{st}$ and $N$, one can
consider the full Born-Infeld dynamics of the D6-brane
\cite{CM97,Gibbons97}, and infer the existence of a D4-brane BIon
state.  One of the main observation from the earlier work on BIons is
the fact that the branes bend in a particular way to preserve the BPS
condition.  It would be very interesting to compare aspects of such a
state to our supergravity configuration valid at large $N$. The fact
that these configurations are BPS should allow some concrete
comparison.

Since the directions shared by the D4-brane and D6-brane are
isometrics, one can easily T-dualize along these directions. This will
give us a localized intersection of D1-brane on D3-brane.  Near
horizon geometry of such a configuration will be of great interest
since the gauge theory on D3-brane {\em does} decouple in the near
horizon limit, and technique similar to ones employed in
\cite{Balasubramanian:1998de} should allow a direct comparison between
supergravity solution and some aspect of the Prasad-Sommerfield magnetic
monopole. In fact, the T-dualizing along the (1,2,3) direction leads to 
dramatic simplification in the form of the dilaton background
$$ {1 \over g_{st}} e^{\phi} = \sqrt{f}$$
which is closely related to the fact that in the absence of D1-brane,
the dilaton background is constant. Unfortunately, T-duality for
supergravity solution is only capable of recovering solutions smeared
along the dual circle.  Furthermore, decoupling and smearing are
non-commutative operation. Therefore, we can not expect to recover any
information about 3+1 dimensional Prasad-Sommerfield monopole directly
from our supergravity solution.

We also discussed a general class\footnote{Similar idea can be used
for branes intersecting Melvin flux tubes. See \cite{Russo:1998xv}.}
of M5-branes embedding along a curve $(v-aw)^N = b^N$ which admits an
analogous construction.  We showed that in general, these curves
describe a paraboloid in the near-horizon geometry of the D6-brane
tensored with flat 3+1-dimensional worldvolume embedded along the
longitudinal coordinates of the D6-brane. In a certain limit, this
paraboloid degenerates in to a ``string'' like configuration
terminated at the core of the D6-brane.  In a certain sense, the
paraboloid can be interpreted as an NS5-brane and its degeneration
into string as a D4-brane. However, the background NSNS 6-form and RR
5-form do not to match this identification. It is possible that such
an identification is only valid in the asymptotically flat region of
space-time and is inapplicable in the near horizon region.

The results reported in this paper are far from solving the
long-standing problem of the spacetime background for all allowed
brane intersections and junctions.  Nonetheless, it is our hope that
having explicit solution at our disposal, albeit in a very special
case, will provide some useful hint or at least point in the right
direction for future work.

\section*{Acknowledgments}

It is a pleasure to thank Arkady Tseytlin for correspondence and
illuminating discussions. I am also grateful to Ofer Aharony, Per
Berglund, Eric D'Hoker, Sunny Itzhaki, and Shamit Kachru for enlightening
discussions. This work was supported in part by the National Science
Foundation under Grant No. PHY94-07194.

\begingroup\raggedright\endgroup

\end{document}